# SPHERE: Meaningful and Inclusive Sensor-Based Home Healthcare

**Alison Burrows**
SPHERE IRC
Merchant Venturers Building
Woodland Road
Bristol BS8 1UB, UK
alison.burrows@bristol.ac.uk

**Ian Craddock**
SPHERE IRC
Merchant Venturers Building
Woodland Road
Bristol BS8 1UB, UK
ian.craddock@bristol.ac.uk

**Rachael Gooberman-Hill**
Bristol Implant Research
Avon Orthopaedic Centre
Southmead Hospital
Bristol BS10 5NB, UK
r.gooberman-hill@bristol.ac.uk

**David Coyle**
Merchant Venturers Building
Woodland Road
Bristol BS8 1UB, UK
david.coyle@bristol.ac.uk

## Abstract
Given current demographic and health trends, and their economic implications, home healthcare technology has become a fertile area for research and development. Motivated by the need for a radical reform of healthcare provision, SPHERE is a large-scale Interdisciplinary Research Collaboration that aims to develop home sensor systems to monitor people's health and wellbeing in the home. This paper outlines the unique circumstances of designing healthcare technology for the home environment, with a particular focus on how to ensure future systems are meaningful to and desirable for the intended users.

## Introduction
Current demographic trends mean healthcare expenditure is rising rapidly. In many parts of the world, including the UK, this is having a noticeable impact on everyday quality of life. As economies stretch and reform in an attempt to accommodate the ultimately unsustainable cost of healthcare, smart home technology presents one promising solution. This has fuelled a policy vision of informed and empowered patients, who play an active role in their healthcare management [1].

It is within this context that the SPHERE (Sensor Platform for HEalthcare in a Residential Environment) Interdisciplinary Research Collaboration (IRC) was established, with the overarching goal of developing practical technology to target a range of health concerns. SPHERE aims to develop new systems that are clinically effective and have the potential for wide spread deployment. Crucially, the IRC aims to combine data-fusion and pattern-recognition from a common platform of currently available sensors in the home environment, rather than to develop sensor technology tailored to specific conditions. Examples of sensors that may be incorporated in SPHERE include video monitoring, environment sensing, and on-body sensing. The central hypothesis is that deviations from people's established pattern of behaviour in their own home have an unexploited diagnostic value.

There is currently a large number of EU and national research projects in the field of Ambient Assisted Living (AAL), partly funded by the Ambient Assisted Living Joint Research and Innovation Programme [2]. These projects have the specific focus of enhancing the quality of life of older people and promoting healthy ageing. In contrast, SPHERE does not target one particular sector of the population. It envisages a system to be used in a household environment and, therefore, to be used by multiple users with a range of needs and abilities. Moreover, this system will take into consideration the constraints and requirements of the healthcare providers, who constitute an equally important stakeholder group.

**Research challenges**
Researchers, developers and policy makers extoll the virtues of various forms of technology interventions to support healthcare, including telehealth (the use of health tracking tools and non-continuous monitoring devices for medical care or treatment purposes), telecare (the use of home-based remote monitoring and assisted living technologies for social care) and AAL. However, despite the availability, affordability and need for these systems, they have not seen widespread adoption. Often these innovations are subject to refusal, intermittent use, misunderstanding, target driven installation, misuse, adaptation, creative use, customization and supplementation [3]. While these situations do not necessarily mean the technology has failed, they do provide evidence of a sociotechnical gap that has yet to be addressed by design.

*Houses versus homes*
An unescapable but underexplored fact about these systems is that they must be designed to be deployed not simply in houses, but in homes. This distinction is crucial, because it takes into account an understanding of the social meaning and value attributed to homes by the people who live in them [4]. The process of making a house a home is defined by a personal appropriation of the space, through physical and often ongoing modifications. Appropriation is also present in human-computer interaction, for instance in the form of hacks and mashups. The transition from technology-as-designed to technology-in-use is mediated through a process of appropriation, which can either reinforce the appropriation-use cycle or eventually lead to disappropriation [5].

There are many open questions on how this appropriation process may be altered or constrained in the case of home-based sensor systems. In particular there is a danger that device appropriation may be constrained if systems focus too rigidly on passive monitoring, which has the potential to hinder user adaptation and the construction of meaning.

*Compliance versus empowerment*
Smart home systems are traditionally enabled by the Ambient Intelligence (AmI) paradigm, which seeks to use pervasive and unobtrusive computing to create user-sensitive and adaptive environments [6]. Research into sensor-based healthcare technology has often operated in the objective realm of data collection, processing and transmission, with a view to monitoring people and prompting them to act as necessary. This outlook seems at odds with the aspiration of the empowered user and has begun to be challenged by projects such as ATHENE, which used ethnography and co-production to enable older people to create technology that mattered to them [7]. The laborious quest for proactive computing systems may have been misguided, since it has operated under the assumption that the onus of decision-making lies with the (smart) technology. Likewise, the pursuit of the 'unobtrusive' may prove to be a false friend, as we are surrounded by countless examples of technology borne with pride as extensions of self, lifestyle and aspirations. Perhaps, as argued by Rogers [8], it is time to steer away from Weiser's vision of calm computing and towards the more realistic goal of using technology ecologies to augment people's ability to learn, decide and perform.

*Single versus multiple user groups*
Research in the home environment poses several challenges, not least of which owing to the presence of multiple users. A home is populated by individuals with different personal characteristics that impact upon technology adoption and use. Specifically, technology adoption is determined by a combination of socio-demographic factors, attitudinal variables, and cognitive abilities [9]. Understanding this user variability is central to meeting the SPHERE objective of producing an inclusive output, which is defined as being functional, usable, desirable, and viable [10]. It is envisaged that taking a broad approach will uncover a plethora of technology and healthcare practices in the home, thus revealing fruitful design opportunities for further exploration.

## Methodology

The SPHERE IRC is still in its infancy, operating within a five year timescale. It intends to establish early and sustained user involvement with a view to producing meaningful technological outputs. In a first phase, we will conduct qualitative ethnographic studies with 15 to 20 households. To achieve this, we will draw on our collaboration with the Knowle West Media Centre in Bristol (UK), which has a recognized track record of community-based technology pilots and is a member of the European Network of Living Labs. Other participants will be recruited through the National Health Service (NHS) and a long-term health research project called the Avon Longitudinal Study of Parents and Children (ALSPAC). Our user-centred design framework will further involve collaboration with clinicians and other healthcare professionals, who are key informants of the therapeutic impact of any system [11].

The SPHERE design research methodology will reflect the need for a contextual understanding of technology use, healthcare practices and related behaviours. Conducting research in the home environment evidently entails practical and ethical challenges. Accordingly, this IRC seeks to use participatory techniques that empower participants in the research process such as:

- Cultural probes [12], which are packages containing various evocative artefacts that invite or provoke people to capture glimpses of their lives – these will be deployed early in the user studies;

- Technology probes [13], which involve installing prototypes into a real use context to gather information about their users and inspire ideas for new technology – these will be informed by findings from the cultural probes and enabled by the expertise of computer science researchers within SPHERE;

- Technology tours [14], in which participants show the researchers around their home and discuss the technology present in each room.

Overall, it is anticipated that SPHERE will contribute to knowledge about interaction with domestic healthcare technology, alongside producing user-sensitive tangible outputs.

## Acknowledgements

The authors gratefully acknowledge funding for the SPHERE IRC from the UK Engineering and Physical Sciences Research Council (EPSRC).